# PIN-Diode-Controlled X-/Ku-Band Frequency Reconfiguration of a CPW-Fed Quasi-Yagi Antenna

Huijuan Niu, Mengyu Dong, Yuanhao Li, Bin Shao, Qingtao Chen, *Member, IEEE,* Junqiang Zhang, Jiawei Tao, Zhenzhong Chen, Gaoming Xu, Gongqing Li, Yongqing Huang, *Member, IEEE,* and Chenglin Bai, *Member, IEEE*

***Abstract*—A PIN-diode-reconfigurable quasi-Yagi antenna fed by a coplanar waveguide (CPW) is presented for X-/Ku-band frequency reconfiguration. PIN diodes are strategically introduced between the driven and parasitic elements to control the effective surface-current paths and thereby modify the operating frequency characteristics. By optimizing the diode locations, number, and switching states, two distinct operating modes are achieved. In Mode I, dual-band operation covering the X- and Ku-bands is obtained, with simulated impedance bandwidths of 9–11 GHz and 12.2–18.8 GHz. In Mode II, a simulated continuous impedance bandwidth from 10 to 14.2 GHz is achieved. The corresponding simulated peak gains are approximately 8 and 7 dBi for Mode I and Mode II, respectively. A prototype is fabricated and experimentally characterized in terms of its S-parameters, and the measured results confirm the frequency-reconfiguration behavior of the proposed antenna. The effects of the number and placement of the PIN diodes on the impedance characteristics are further investigated to clarify their roles in frequency reconfiguration and design complexity. The proposed antenna provides a compact CPW-fed quasi-Yagi configuration with PIN-controlled X-/Ku-band frequency reconfiguration.**

This work was funded by the natural science fund project of Shandong Province (ZR2022MF305, ZR2022MF253). The undergraduate teaching reform research project of the Education Department of Shandong Province (Z2023010). The Science and Technology SMES Innovation Ability Improvement Project of Shandong Province (2022TSGC2570). *(Corresponding author: Qingtao Chen).*

Huijuan Niu, Mengyu Dong, Yuanhao Li, Junqiang Zhang, Jiawei Tao, and Chenglin Bai are with the School of Physics Science and Information Technology, Liaocheng University, Liaocheng 252000, China, and also with the Liaocheng Key Laboratory of Industrial-Internet Research and Application, Liaocheng 252000, China (e-mail: supernhj@lcu.edu.cn).

Bin Shao, Zhenzhong Chen, and Gaoming Xu are with the Faculty of Electrical Engineering and Computer Science, Ningbo University, Ningbo, Zhejiang 315211, China.

Qingtao Chen is with the Department of Electrical and Computer Engineering, McGill University, Montreal, QC H3A 0E9, Canada (e-mail: qingtao.chen@mcgill.ca).

Gongqing Li is with the Space Engineering University, Beijing 101416, China.

Yongqing Huang is with the State Key Laboratory of Information Photonics and Optical Communications, Beijing University of Posts and Telecommunications, Beijing 100876, P. R. China.

## I. Introduction

ANTENNAS play an important role in RF front ends by converting guided electromagnetic signals into free-space radiation. They are also widely used in optoelectronic systems, where high-speed, high-power, and high responsivity photodetectors [1]–[9] and photonic integrated or hybrid electronic–photonic circuits provide a platform for generating microwave and millimeter-wave signals for communication and sensing applications [10]–[14]. The antenna performance, including bandwidth, efficiency, gain, and radiation characteristics, has a direct impact on the overall system performance [15]–[18]. Among various antenna configurations, end-fire antennas such as Yagi–Uda antennas offer high directivity and front-to-back ratio through the interaction of driven elements, reflectors, and directors [19]–[21]. Printed quasi-Yagi antennas retain these advantages while providing a compact profile and compatibility with microwave and millimeter-wave integrated circuits [22]–[25]. However, conventional Yagi antennas usually operate over a limited bandwidth with fixed resonant frequencies, which has motivated the development of frequency-reconfigurable Yagi antennas for multiband applications [26], [27].

To realize frequency reconfiguration in Yagi-type antennas, existing approaches can be broadly classified into three categories. First, PIN diodes or RF micro-electro-mechanical-system (MEMS) switches are used to connect or isolate resonant sections, thereby producing predefined frequency states [28]–[32]. Independently tunable bandwidths have also been demonstrated by employing PIN diodes to modify the effective resonator length [33]. Second, continuously tunable components or materials, such as varactor diodes and liquid crystals, adjust the effective electrical length of the antenna elements [27], [31], [34]. Although this approach enables continuous frequency tuning, it requires precise analog biasing and is susceptible to nonlinearity, loss, temperature variations, and limited power handling. Third, frequency reconfiguration can be achieved by switching segmented or parasitic elements to modify the current distribution and the functions of the driven, reflector, and director elements [30], [31]. This approach has also been applied to quasi-Yagi antennas for dual-band frequency reconfiguration [35]. Despite these advances, broadband reconfigurable Yagi-type antennas remain relatively limited.

This work presents a coplanar-waveguide (CPW)-fed frequency-reconfigurable quasi-Yagi antenna for X- and Ku-band applications. PIN diodes inserted between the feedline and the active and parasitic elements reconfigure the surface-current paths and interchange the functions of selected elements. In State 1, the driven element is excited, while a parasitic patch introduces an additional resonance to broaden the operating bandwidth. In State 2, the parasitic patch becomes the main radiator and the original driven element functions as a director, shifting the operating band. The uniplanar CPW feed facilitates impedance matching and bias integration [22]. The proposed antenna achieves dual-state frequency reconfiguration with broadband X-/Ku-band operation and directional radiation. The influence of the number and distribution of PIN diodes is also investigated. Simulated and measured results are in good agreement, demonstrating the effectiveness of the proposed design.

## II. Quasi-Yagi Antenna Design

### A. Antenna Geometry

A typical Yagi–Uda antenna consists of a fed driven element, a reflector, and one or more directors, in which the parasitic elements are excited through mutual electromagnetic coupling [22], [25]. Building on the conventional Yagi–Uda configuration, in which the driven element, reflector, and directors have typical lengths of approximately $0.5\lambda$, $0.6\lambda$, and $0.45\lambda$, respectively [24], a frequency-reconfigurable quasi-Yagi antenna for X- and Ku-band applications is proposed with the PIN diodes shown in red, as illustrated in Fig. 1. The antenna comprises an upper radiating section and a lower coplanar waveguide (CPW) feed network fabricated on a Rogers RT/duroid®6010LM substrate with a relative dielectric constant of 10.7 and a thickness of 0.635 mm.

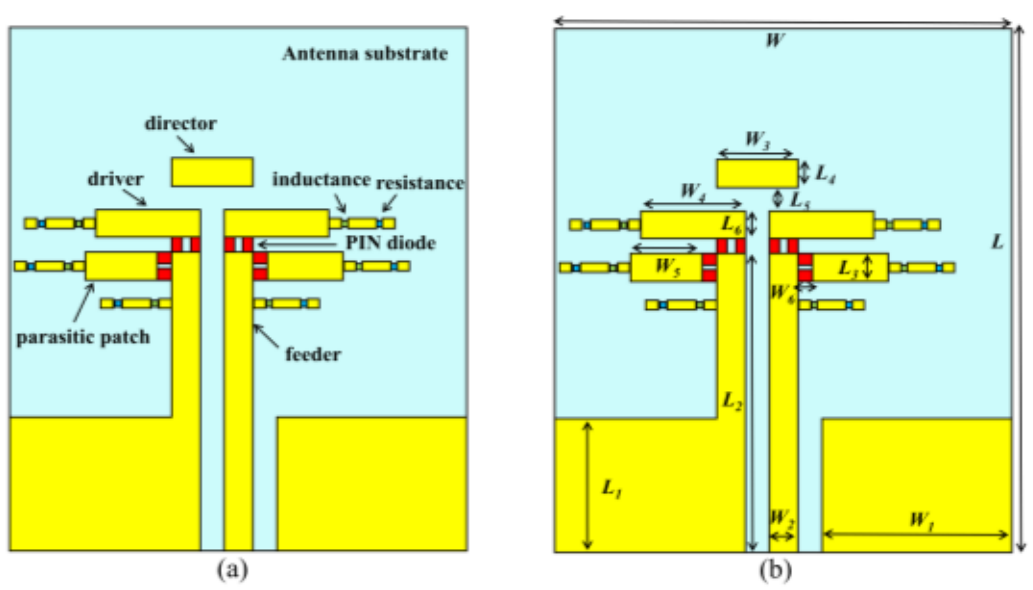

Fig. 1. The proposed reconfigurable Yagi antenna: (a) Antenna schematic view, and (b) Detailed geometric dimensions in mm scale: $W$=19, $W_1$ = 8.3, $W_2$ = 1.3, $W_3$ = 3.05, $W_4$ = 3.7, $W_5$ = 2.1, $L$ = 23, $L_1$ = 4.5, $L_2$ = 13.35, $L_3$ = 1, $L_4$ = 1, $L_5$ = 1.25, and $L_6$ = 1.

### B. Operating Principle

According to microstrip antenna theory, introducing a parasitic patch adjacent to the driven element generates an additional resonance, thereby broadening the operating bandwidth [33]. In the proposed design, a rectangular parasitic patch is placed above the driven element to produce a dual-resonance response. Frequency reconfiguration is realized by incorporating MADP-000907-14020 PIN diodes between the feedline and the active and parasitic elements, allowing the current paths and the functions of the radiating elements to be reconfigured. The PIN diodes exhibit a typical capacitance of 0.30 pF in the OFF state and a resistance of 7.8 Ω under forward bias [36]. Based on the current distribution, the bias network is arranged in low-current regions at the ends of the radiating elements and along the feedline, minimizing its influence on antenna performance without requiring additional DC-blocking capacitors [14], [18], [35].

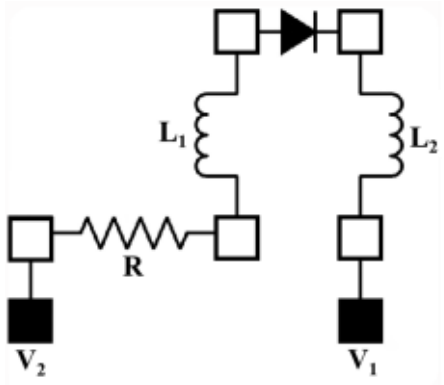

Fig. 2. Schematic of the DC biasing circuit for PIN diode.

The bias circuitry employs L-05B10NJV6T inductors (10 nH), which isolate the DC bias path from the high-frequency signal. To protect the circuit from excessive current, a 300 Ω resistor is included in the bias network. As illustrated in Fig. 2, the diode switching states are controlled by varying the supply voltages $V_1$ and $V_2$. When $V_2$=3 V and $V_1$=0 V, the diode is ON. When $V_2$=0 V, the diode is OFF. This configuration enables precise control of the PIN diode's operational mode, allowing the antenna to switch between two distinct resonance modes.

The operating principle of this frequency-reconfigurable antenna relies on strategically placing PIN diodes between the feedline and the active parasitic elements, as well as between the feedline and the parasitic patches. In State 1, when the PIN diodes between the feedline and the driven element are forward-biased, the antenna operates according to microstrip antenna theory. In this case, the parasitic patch functions to enhance the bandwidth. In State 2, when the diode between the feedline and the parasitic patch is forward-biased, the parasitic patch in State 1 becomes the active radiating element, while the driven element in State 1 transitions into a director in State 2. By controlling the ON/OFF states of PIN diodes, the current distribution and radiation behavior of the Antenna can be dynamically altered, thereby enabling frequency reconfigurability [13], [15].

## III. Simulated and Measured Results

### A. Simulation Results for the Reconfigurable Antenna

The proposed frequency-reconfigurable Yagi antenna was designed and optimized using full-wave electromagnetic simulations. During the simulations, the ON state (Mode I) of the PIN diode was modeled by a 7.8 Ω resistor, whereas the OFF state (Mode II) was modeled by a 0.30 pF capacitor. The simulated $S$-parameter results for the two operating modes are shown in Fig. 3.

Simulation results show that by switching between two operating states, this antenna essentially achieves full coverage of the X and Ku bands. In State 1 (when the diode between the feedline and the driven element is conducting, Fig.3(a)), the parasitic patch effectively broadens the frequency band, resulting in bandwidths of 9-11 GHz and 12.2-18.8 GHz, representing a significant 215% increase over the primary antenna (Fig. 3(a)), with an average gain of 6.2 dBi within this band (and a maximum gain of 7.9 dBi at 17.2 GHz). When switched to State 2 (with the diode between the feedline and the parasitic patch conducting, Fig.3(b)), the antenna bandwidth shifts to 10-14.2 GHz, achieving an average gain of 4.5 dBi within this frequency range (with a maximum gain of 6.8 dBi at 12.2 GHz).

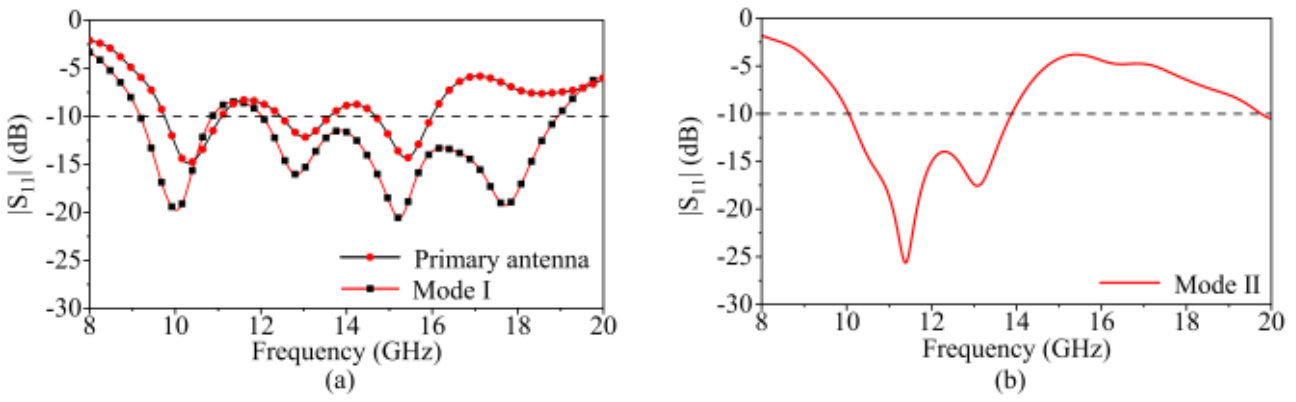

Fig. 3. Reflection coefficients of a frequency-reconfigurable Yagi antenna. (a) Mode I and (b) Mode II.

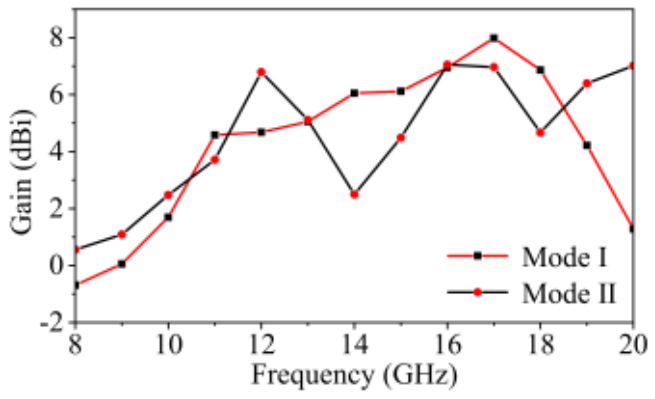

Fig. 4. Simulation results of antenna gain at Mode I and Mode II.

Figure 4 illustrates the antenna gain in both states. In State 1, with $S_{11}\leq$ -10 dB, the gain corresponding to the bandwidth of 12.2 GHz–18.8 GHz ranges from 4.5 dBi to 4.8 dBi, reaching a maximum of 7.9 dBi at 17.2 GHz. The average gain within the operational bandwidth is 6.2 dBi. In State 2, with $S_{11}\leq$-10 dB, the gain over the 10 GHz–14.2 GHz bandwidth varies from 2.5 dBi to 3 dBi, with a peak gain of 6.8 dBi at 12.2 GHz, and an average gain of 4.5 dBi throughout the working frequency range.

### *B. Effect of Diode Number on Antenna Performance*

The antenna reconfigurability depends on both the switching states and the number of PIN diodes. Therefore, the effect of diode number on antenna performance is investigated here. Figure 5 illustrates four antenna configurations with different PIN-diode arrangements: Antenna I with four diodes, Antenna II and Antenna III with two six-diode configurations, and Antenna IV with eight diodes.

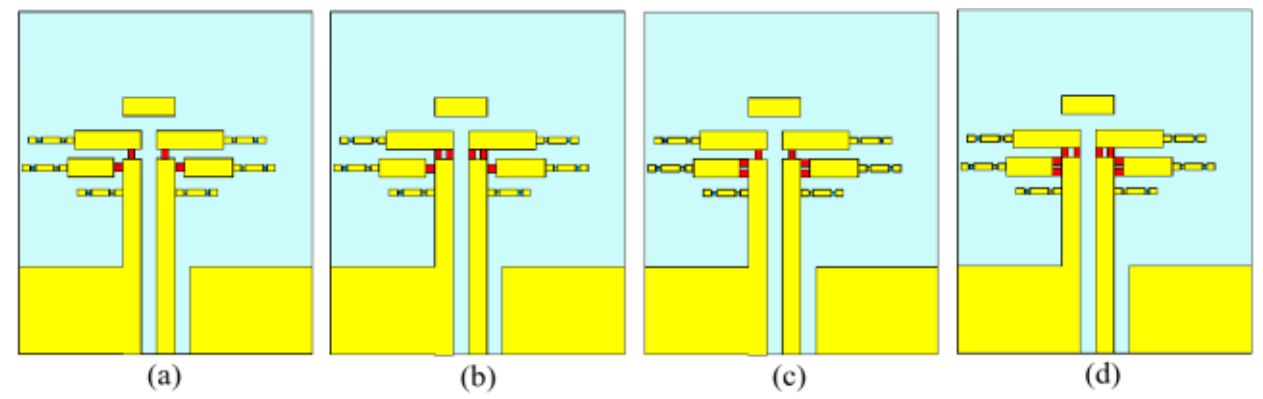

Fig. 5. Schematic diagrams of the reconfigurable antenna configurations with different PIN-diode arrangements: (a) Antenna I: 4 PIN diodes, (b) Antenna II: 6 PIN diodes, (c) Antenna III: 6 PIN diodes, and (d) Antenna IV: 8 PIN diodes. The number and distribution of PIN diodes vary among the driven elements, parasitic elements, and feedline.

The reflection coefficients and gains of the four antenna configurations are shown in Figs. 6 and 7. In Mode 1 (Fig. 6(a)), Antennas I and III exhibit similar −10 dB impedance bandwidths of 8.2–10.3 GHz and 11.8–17.8 GHz, respectively, whereas Antennas II and IV achieve bandwidths of 9–11 GHz and 12.2–18.8 GHz, respectively. In Mode 2 (Fig. 6(b)), Antennas I and II exhibit similar −10 dB impedance bandwidths of 9.5–13.5 GHz, while Antennas III and IV achieve bandwidths of 10–14.2 GHz. Regarding the gain performance (Fig. 7), in Mode 1 (Fig. 7(a)), Antennas I and III achieve a maximum gain of 7.8 dBi at 17 GHz, while Antennas II and IV reach 7.6 dBi. However, Antennas II and IV exhibit an average gain approximately 1 dB higher than that of Antennas I and III over the 11–15.5 GHz band. In Mode 2 (Fig. 7(b)), Antennas I and II achieve maximum gains of 5 dBi and 4.5 dBi at 12 GHz, respectively, whereas Antennas III and IV both reach 6.8 dBi. Moreover, Antennas III and IV provide an average gain improvement of approximately 2 dB compared with Antennas I and II over the 11–13.5 GHz band.

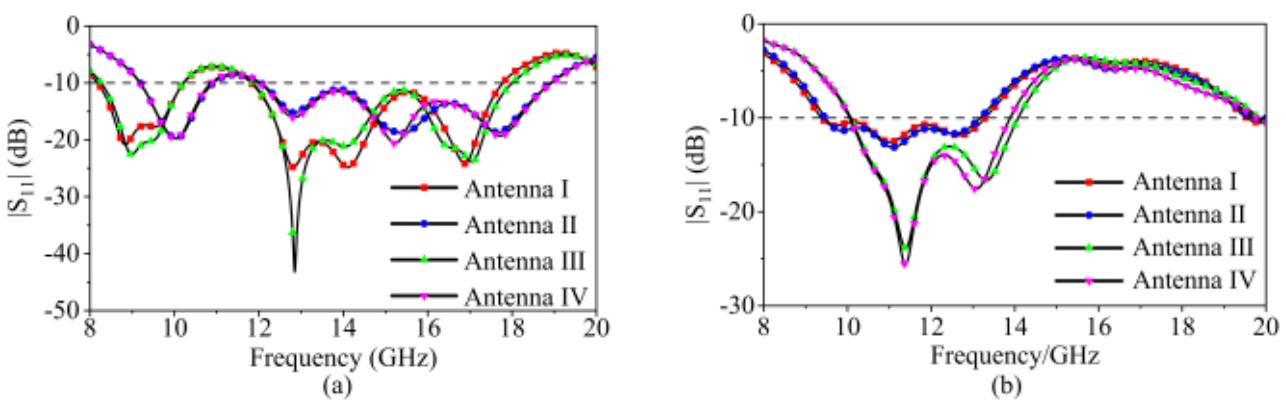

Fig. 6. Reflection coefficients of four antennas. (a) Mode I and (b) Mode II.

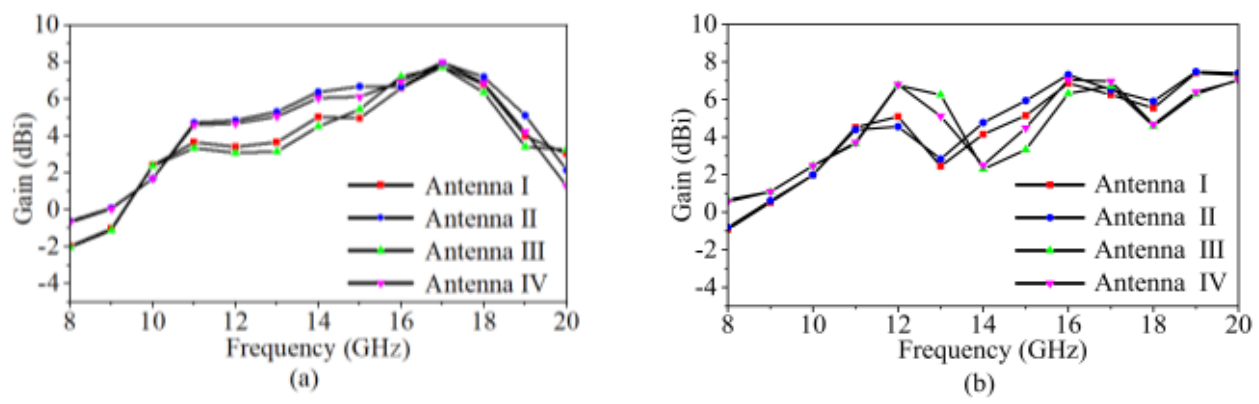

Fig. 7. Gain of the four antennas. (a) Mode I and (b) Mode II.

In conclusion, varying the number of PIN diodes on one side of the antenna does not compromise the performance of the other side. Moreover, all four antenna configurations maintain favorable directional radiation characteristics, as verified by their radiation patterns.

### *C. Effect of $W_5$ on Antenna Performance*

Based on the previous simulation results, the configuration with eight PIN diodes (Antenna IV) is selected as the optimal design and reference configuration for further study. The effects of $W_5$ on the reflection coefficient and gain of Antenna IV are investigated, as shown in Figs. 8 and 9, respectively.

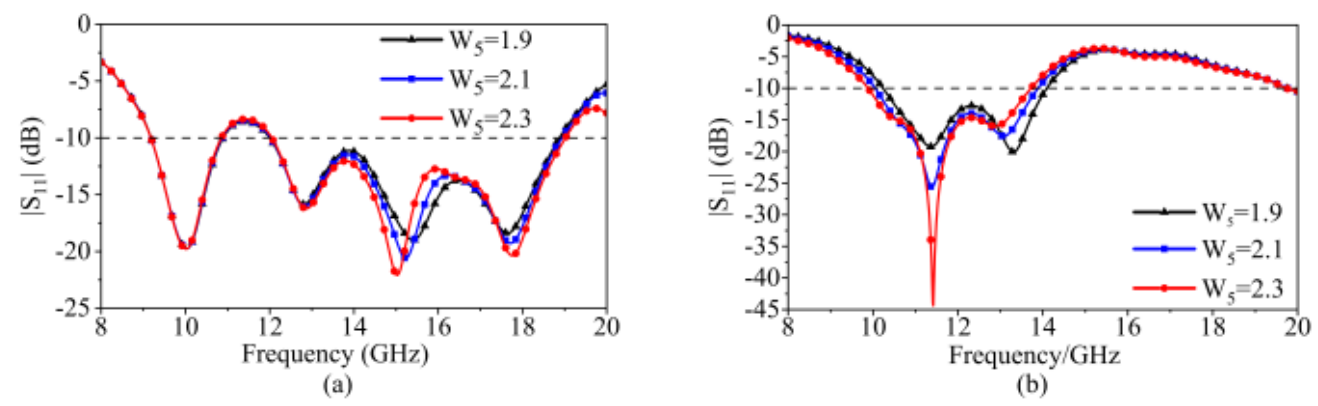

Fig. 8. Effect of $W_5$ on Antenna IV reflection coefficient in (a) State 1 and (b) State 2.

As shown in Figs. 8 and 9, increasing $W_5$ slightly broadens the bandwidth in State 1 and shifts the operating band toward lower frequencies in State 2, while affecting the resonance and gain characteristics in both states. The analysis of $W_5$ shows that trade-off among multiple performance metrics is required, and $W_5$ = 2.1 mm is selected as the final design value.

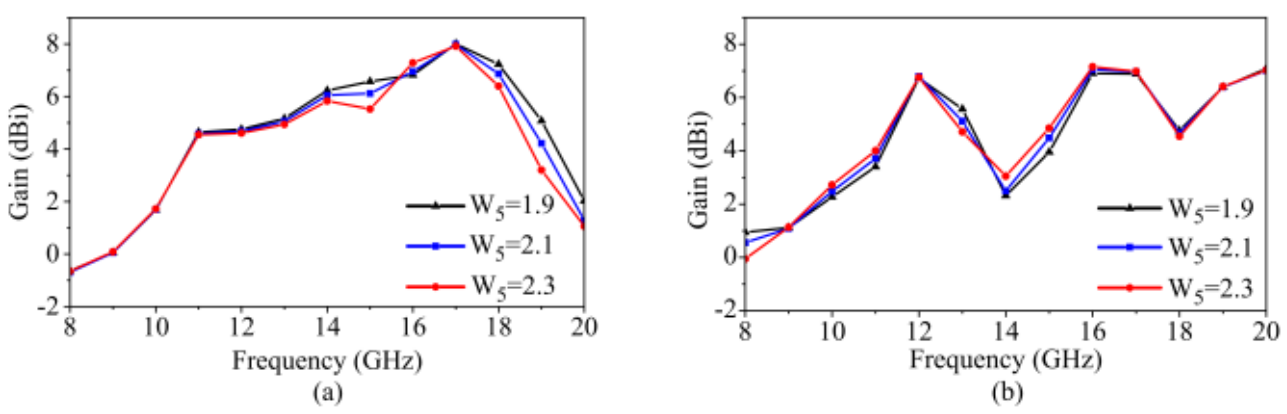

Fig. 9. Effect of $W_5$ on Antenna IV gain in (a) State 1 and (b) State 2.

### *D. Measured Results*

Photographs of the fabricated antenna prototypes are shown in Fig. 10. The antenna prototypes are fabricated on Rogers RT/duroid 6010LM substrates with a thickness of 0.635 mm and a relative permittivity of 10.7. The 0.035-mm-thick copper layer consists of the driven element, parasitic elements, director, PIN diodes, and bias network. The antenna prototypes are connected to the measurement system through SMA connectors. The PIN diodes, inductors, and resistors are soldered at the designated positions, and the bias networks are connected using wires of equal length. The diode ON/OFF states are controlled by adjusting the bias voltage.

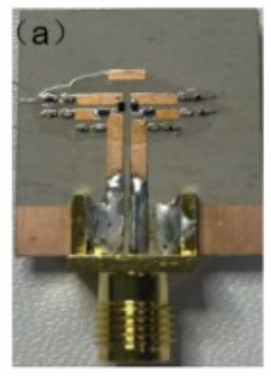

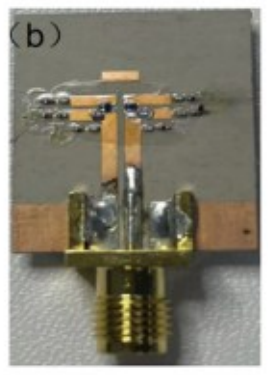

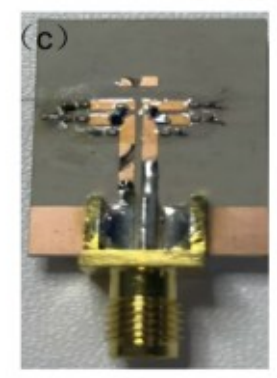

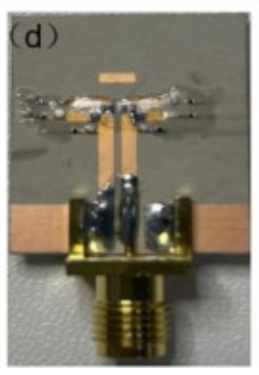

Fig. 10. Photographs of the fabricated antennas. (a) Antenna I, (b) Antenna II, (c) Antenna III, and (d) Antenna IV.

Due to measurement limitations, only the reflection coefficients of the proposed frequency-reconfigurable Yagi antenna were measured up to 18 GHz. The antenna was switched between the two operating states by adjusting the bias voltage: State 1 was achieved by applying 3 V to the feedline and 0 V to the driven element, whereas State 2 was achieved by applying 3 V to the parasitic patch and 0 V to the feedline.

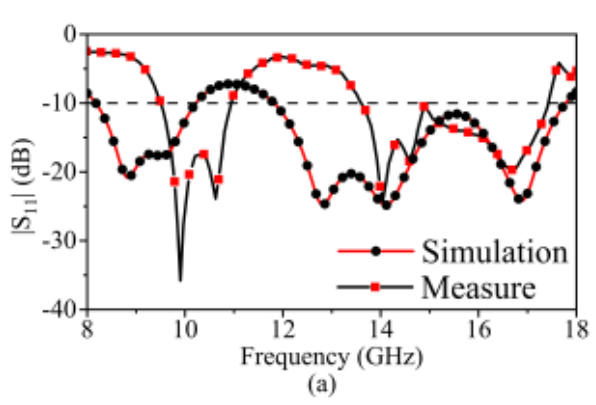

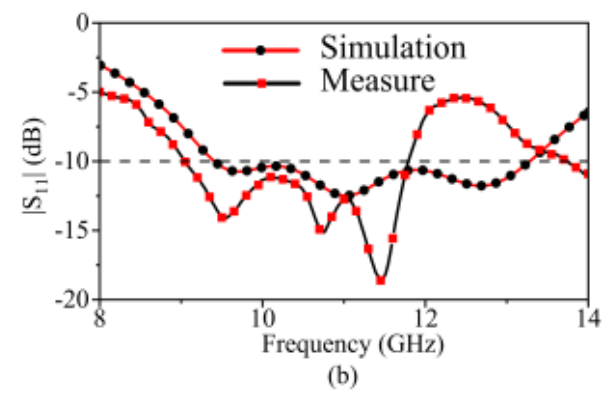

Fig. 11. Comparison of the simulated and measured reflection coefficients ($S_{11}$) of Antenna I. (a) State 1 and (b) State 2.

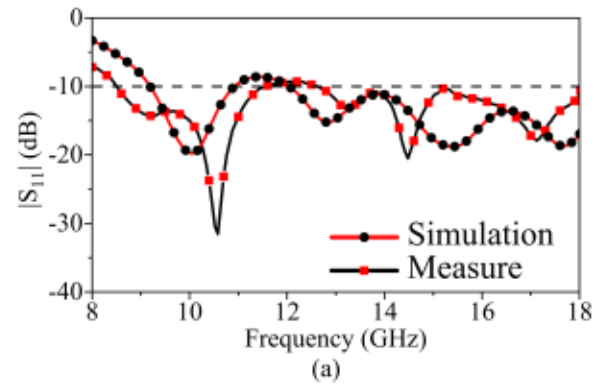

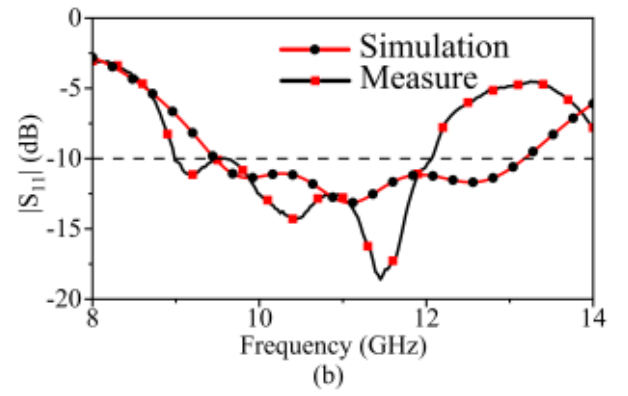

Fig. 12. Comparison of the simulated and measured reflection coefficients ($S_{11}$) of Antenna II. (a) State 1 and (b) State 2.

Figures 11–14 compare the simulated and measured reflection coefficients ($S_{11}$) of Antennas I-IV in the two switching states. In State 1, all four antenna prototypes exhibit dual-band operation. As shown in Fig. 11(a), Antenna I operates over 9.50–10.95 GHz and 13.65–17.72 GHz. Similarly, Antennas II, III, and IV cover 8.50–11.55 GHz and 12.80–18.30 GHz (Fig. 12(a)), 8.30–11.30 GHz and 12.60–17.60 GHz (Fig. 13(a)), and 8.30–11.25 GHz and 12.50–18.10 GHz (Fig. 14(a)), respectively. In State 2, each prototype exhibits a single lower-frequency band. The corresponding bandwidths are 9.05–11.80 GHz for Antenna I (Fig. 11(b)), 8.90–12.00 GHz for Antenna II (Fig. 12(b)), 9.30–11.40 GHz for Antenna III (Fig. 13(b)), and 8.80–11.80 GHz for Antenna IV (Fig. 14(b)). The measured results verify the intended state-dependent transition from dual-band to lower-frequency single-band operation.

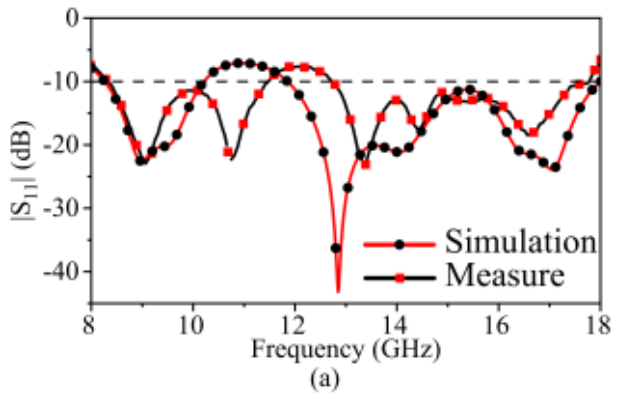

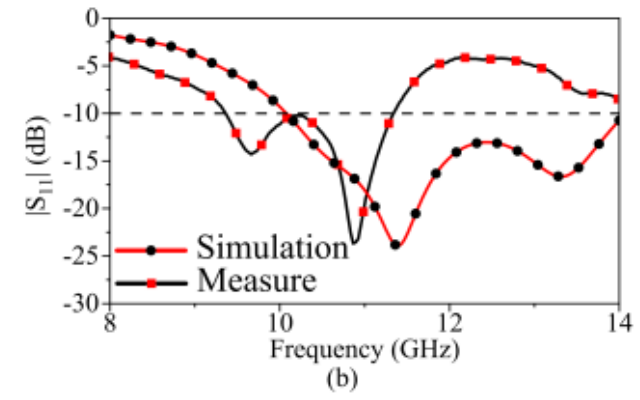

Fig. 13. Comparison of the simulated and measured reflection coefficients ($S_{11}$) of Antenna III. (a) State 1 and (b) State 2.

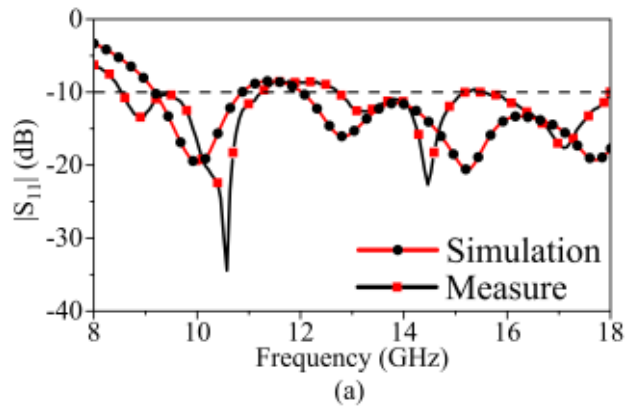

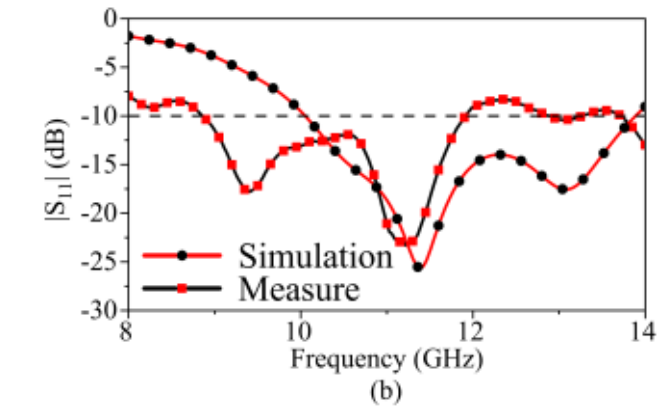

Fig. 14. Comparison of the simulated and measured reflection coefficients ($S_{11}$) of Antenna IV. (a) State 1 and (b) State 2.

Overall, although the measured center frequencies deviate slightly from the simulated results, the measured and simulated curves exhibit consistent trends. The discrepancies can be attributed to unavoidable fabrication tolerance, environmental interference during the measurements, and the influence of the solder joints used to connect the PIN diodes to the DC bias network, which alter the surface resistance of the antenna. In addition, the close spacing between adjacent PIN diodes introduces mutual interactions. These factors collectively may have caused the observed shift in the reflection coefficient curves.

## IV. CONCLUSION

This paper presents a frequency-reconfigurable CPW-fed quasi-Yagi antenna for X- and Ku-band applications. Frequency reconfiguration is achieved by incorporating PIN diodes between the driven element and the parasitic elements, allowing the radiation state and resonant frequency of the antenna to be controlled through RF switching. The proposed antenna operates in two states, covering 9–11 GHz and 12.2–18.8 GHz in State 1, and 10–14.2 GHz in State 2. The average realized gains over the corresponding operating bands are 6.5 dBi and 4.2 dBi, respectively. Furthermore, four antenna configurations with different numbers and distributions of PIN diodes are designed and investigated. The results show that all four configurations maintain good directional radiation characteristics, demonstrating their potential for reconfigurable X-/Ku-band wireless communication and radar applications.